\documentclass[preprint,amsmath,amssymb,nofootinbib]{revtex4}
\usepackage[utf8]{inputenc}
\usepackage{rotating}
\usepackage{hyperref}
\usepackage{array}
\usepackage{subfig}
\usepackage{float}
\usepackage{graphicx}
\usepackage{slashed}
\usepackage{color}
\usepackage{ctable}
\definecolor{orange}{rgb}{1,0.5,0}
\usepackage[T1]{fontenc}

\begin{document}

\title{Lepton collider probes for Majorana neutrino effective interactions.}

\author{Gabriel Zapata}
\affiliation{Instituto de F\'{\i}sica de Mar del Plata (IFIMAR)\\ CONICET, UNMDP\\
Funes 3350, (7600) Mar del Plata, Argentina} 


\author{Tom\'as Urruzola}
\affiliation{Instituto de F\'{\i}sica, Facultad Ingenier\'ia,
 Universidad de la Rep\'ublica \\ Julio Herrera y Reissig,(11300) 
Montevideo, Uruguay.}

\author{Oscar A. Sampayo}
\email{sampayo@mdp.edu.ar}
\affiliation{Instituto de F\'{\i}sica de Mar del Plata (IFIMAR)\\ CONICET, UNMDP\\ Departamento de F\'{\i}sica,
Universidad Nacional de Mar del Plata \\
Funes 3350, (7600) Mar del Plata, Argentina}

\author{Luc\'{\i}a Duarte}
\email{lucia@fisica.edu.uy}
\affiliation{Instituto de F\'{\i}sica, Facultad de Ciencias,
 Universidad de la Rep\'ublica \\ Igu\'a 4225,(11400) 
Montevideo, Uruguay.}

\begin{abstract}

The extension of the standard model with new high-scale weakly coupled physics involving right-handed neutrinos in an effective field theory framework (SMNEFT) allows for a systematic study of heavy neutrinos phenomenology in current and future experiments. We exploit the outstanding angular resolution in future lepton colliders to study the sensitivity of forward-backward asymmetries to discover the possible single production of heavy Majorana neutrinos via $e^{+}e^{-} \to N \nu$, followed by a purely leptonic decay $N \to \mu^{-} \mu^{+} \nu$ or a semi-leptonic decay $N \to \mu^{-} \mathrm{j} \mathrm{j} $, for masses $m_N > 50$ GeV. In this regime, we consider the $N$ production and decays to be dominated by scalar and vectorial four-fermion $d=6$ single $N_R$ operators. This is an alternative analysis to searches using displaced vertices and fat jets, in a higher mass regime, where the $N$ is short-lived but can be found by the angular distribution of its decay products. We find that a forward-backward asymmetry between the final muons in the pure leptonic decay mode provides a sensitivity up to 12$\sigma$ for $m_N=100$ GeV, for effective couplings $\alpha=0.2$ and new physics scale $\Lambda=1$ TeV. In the case of the semi-leptonic decay, we can compare the final muon and higher $p_T$ jet flight directions, again finding up to 12$\sigma$ sensitivity to the effective signals.

\end{abstract}

\maketitle

\section{Introduction.}{\label{intro}}

The simplest Standard Model (SM) extensions that can account for light neutrino masses, and hence the neutrino oscillations phenomena, predict the existence of sterile right-handed neutrinos $N_R$ with Majorana mass terms, as the Type I \cite{Minkowski:1977sc, Mohapatra:1979ia, Yanagida:1980xy, GellMann:1980vs, Schechter:1980gr} and also the linear and inverse seesaw mechanisms. These are weakly coupled fields possibly at the electroweak (EW) scale and their interactions with the SM particles, beyond the seesaw mixing with active neutrino states, could be described by an effective field theory (EFT) including them as low energy degrees of freedom: the Standard Model Effective Field Theory framework extended with right-handed Neutrinos (SMNEFT) \footnote{Also called $N_R$SMEFT and $\nu_R$SMEFT in the literature.}, with operators known up to dimension $d=9$ \cite{delAguila:2008ir, Aparici:2009fh, Liao:2016qyd, Bhattacharya:2015vja, Li:2021tsq}. On the other hand, approaches like the so called neutrino non standard interactions (NSI) and general neutrino interactions (GNI) are incorporating right-handed neutrinos to the Standard Model effective field theory (SMEFT) \cite{Bischer:2019ttk, Han:2020pff, Terol-Calvo:2019vck, Escrihuela:2021mud} considering their Majorana and/or Dirac nature.

The extensions of the Type I seesaw renormalizable lagrangian with effective interactions of higher dimension for the right-handed neutrinos are getting attention, since the naive seesaw heavy $N$ production rates and decay widths can be challenged by the effective interactions, leading to a variety of signals that can be studied at the LHC and different colliders \cite{Caputo:2017pit,Barducci:2022hll, Barducci:2020icf, Alcaide:2019pnf, Butterworth:2019iff, Biekotter:2020tbd, DeVries:2020jbs,Cirigliano:2021peb, Zhou:2021ylt, Cottin:2021lzz, Beltran:2021hpq, Jones-Perez:2019plk, Magill:2018jla}.

We will focus here on a simplified scenario with only one heavy Majorana neutrino $N$, considering its interactions to be dominated by new physics at a higher energy scale and parameterized by the $d=6$ SMNEFT operators, and neglect the effect of the renormalizable Yukawa lagrangian term $N L \phi$ that gives place to the heavy-active neutrino mixings $U_{l N}$, as it is strongly constrained not only by the naive seesaw relation $U_{lN}^2 \sim m_{\nu}/M_N \sim 10^{-14}-10^{-10}$ required to account for the light $\nu$ masses \cite{Atre:2009rg}, but also by experimental constraints. Our previous works involved the Majorana $N$ decays \cite{Duarte:2015iba, Duarte:2016miz}, surveyed its production signals in colliders for masses $m_N$ above the EW-scale \cite{Peressutti:2011kx, Peressutti:2014lka, Duarte:2014zea, Duarte:2018xst, Duarte:2018kiv} focusing on lepton number violating (LNV) processes, and in the order $m_N=1-10$ GeV scale, where its decay is dominated by the $N\to \nu \gamma$ channel \cite{Duarte:2016caz}, and can be produced in the decays of B mesons \cite{Duarte:2019rzs, Duarte:2020vgj}. 

The search for EW scale massive neutrino states $N$ is among the goals of future $e^{+} e^{-}$ colliders, as they provide a clean environment with lower SM backgrounds compared to an hadronic machine. The literature regarding the use of lepton colliders in past, existing and proposed experiments like the linear ILC \cite{Behnke:2013xla} or circular colliders like the FCC-ee \cite{FCC:2018evy} and the CEPC \cite{CEPCStudyGroup:2018ghi} to study the production of heavy neutrinos is very extensive \cite{Barducci:2020ncz, Barducci:2022hll, Barducci:2020icf, Hernandez:2018cgc, Das:2018usr, Chakraborty:2018khw, Zhang:2018rtr, Liao:2017jiz, Yue:2017mmi, Biswal:2017nfl, Antusch:2016ejd, Antusch:2015rma, Antusch:2015mia, Banerjee:2015gca, Blondel:2014bra}.

In this article we want to exploit the remarkable angular resolution that will be available for the detection of final leptons in future lepton colliders to study the sensitivity of forward-backward asymmetries to discover the possible $N$ production in pure leptonic and semi-leptonic final states. Thus, we analyse the single production of heavy Majorana neutrinos via $e^{+}e^{-} \to N \nu$, followed by a purely leptonic decay $N \to \mu^{-} \mu^{+} \nu$ or a semi-leptonic decay $N \to \mu^{-} \mathrm{j} \mathrm{j} $, for masses $m_N > 50$ GeV. Our aim is to present an alternative analysis to searches using displaced vertices \cite{Beltran:2021hpq, Duarte:2016caz} in the lower mass regime, revisiting the study of angular observables in heavy neutral particles phenomenology known since the LEP era \cite{Petcov:1984nf, Bilenky:1986nd, Hofer:1996cs, Cvetic:1998vg}. We upgrade our previous work on this approach, performing a dedicated simulation and analysis of the proposed signals. 

The paper is organized as follows: in Sec.\ref{sec:eff_form} we describe the SMNEFT formalism, its \texttt{FeynRules} implementation in Sec.\ref{sec:feynrulesimplementation} and in Sec.\ref{sec:constraints} we summarize the existing bounds on the effective couplings. In Sec.\ref{sec:Collider analysis} we discuss the signals and the standard model backgrounds features, and present our analysis and results for the leptonic (Sec.\ref{sec:Leptonic}) and semi-leptonic (Sec.\ref{sec:Jets}) channels, closing with a summary in Sec.\ref{sec:summary}.

\section{Effective interactions formalism.}\label{sec:eff_form}

The SM lagrangian is extended with only \emph{one} right-handed neutrino $N_R$ with a Majorana mass term, which gives a massive state $N$ as an observable degree of freedom. The new physics effects are parameterized by a set of effective operators $\mathcal{O}_\mathcal{J}$ constructed with the SM and the $N_R$ fields and satisfying the $SU(2)_L \otimes U(1)_Y$ gauge symmetry \cite{delAguila:2008ir, Liao:2016qyd, Wudka:1999ax}. The effect of these operators is suppressed by inverse powers of the new physics scale $\Lambda$. The total lagrangian\footnote{Note that we do not include the Type-I seesaw lagrangian terms giving the Majorana and Yukawa terms for the sterile neutrinos.} is organized as follows:
\begin{eqnarray}\label{eq:lagrangian}
\mathcal{L}=\mathcal{L}_{SM}+\sum_{d=5}^{\infty}\frac1{\Lambda^{d-4}}\sum_{\mathcal{J}} \alpha_{\mathcal{J}} \mathcal{O}_{\mathcal{J}}^{d}
\end{eqnarray}
where $d$ is the mass dimension of the operator $\mathcal{O}_{\mathcal{J}}^{d}$, $\alpha_{\mathcal{J}}$ are the effective couplings and the sum in $\mathcal{J}$ goes over all independent interactions at a given dimension $d$.

\begin{table}[t]
 \centering
 \begin{tabular}{c l c r c l c r}
 \firsthline \specialrule{.1em}{.05em}{.05em} 
 Operator &  Notation & Type & Coupling  &  Operator &  Notation & Type & Coupling   \\
  \specialrule{.03em}{.03em}{.03em}
 $\;\; (\phi^{\dag}\phi)(\bar L_i N \tilde{\phi}) \;\;$ & $\;\;  \mathcal{O}^{(i)}_{LN\phi} \;\;$   &~S~~& $\alpha^{(i)}_{\phi}$ &  &  &  \\
 \specialrule{.03em}{.03em}{.03em}
 $\;\; i(\phi^T \epsilon D_{\mu}\phi)(\bar N \gamma^{\mu} l_i)\;\;$ & $ \;\;\mathcal{O}^{(i)}_{Nl\phi}\;\;$&~V~~& $\alpha^{(i)}_W$ & $\;\;  i(\phi^{\dag}\overleftrightarrow{D_{\mu}}\phi)(\bar N \gamma^{\mu} N) \;\;$& $\;\; \mathcal{O}_{NN\phi} \;\;$&~V~~& $\alpha_Z$\\ 
\specialrule{.06em}{.05em}{.05em} 
 $\;\;(\bar N \gamma_{\mu} l_i) (\bar d _j \gamma^{\mu} u _j)\;\;$& $\;\; \mathcal{O}^{(i, j)}_{duNl} \;\;$&~V~~& $\alpha^{(i, j)}_{V_0}$ &  $\;\;(\bar N \gamma_{\mu}N) (\bar f_i \gamma^{\mu}f_i) \;\;$& $\;\; \mathcal{O}^{(i)}_{fNN} \;\;$ &~V~~& $\alpha^{(i)}_{V_{f}}$ \\
 \specialrule{.03em}{.03em}{.03em}
 $\;\; (\bar Q _i u _i)(\bar N L_j) \;\;$& $\;\;\mathcal{O}^{(i, j)}_{QuNL}\;\;$ &~S~~& $\alpha^{(i, j)}_{S_1}$ & $ \;\;(\bar L_i N)\epsilon (\bar L_j l_j)\;\; $ & $\;\;\mathcal{O}^{(i, j)}_{LNLl}\;\;$&~S~~& $\alpha^{(i, j)}_{S_0}$\\
 $\;\;(\bar L_i N) \epsilon (\bar Q _j d _j)\;\;$& $\;\;\mathcal{O}^{(i, j)}_{LNQd}\;\;$&~S~~& $\alpha^{(i,j)}_{S_2}$ & \\
 $\;\;(\bar Q _i N)\epsilon (\bar L_j, d _j)\;\;$ & $\;\;\mathcal{O}^{(i, j)}_{QNLd}\;\;$ &~S~~&  $\alpha^{(i, j)}_{S_3}$ & & &\\
  \specialrule{.06em}{.05em}{.05em} 
 $\;\; (\bar L_i \sigma^{\mu\nu} \tau^I N) \tilde \phi W_{\mu\nu}^I \;\;$& $\;\;\mathcal{O}^{(i)}_{NW}\;\;$ & T & $\alpha^{(i)}_{NW}$ &  $\;\;(\bar L_i \sigma^{\mu\nu} N) \tilde \phi B_{\mu\nu}\;\;$& $\;\;\mathcal{O}^{(i)}_{NB}\;\;$ & T &$\alpha^{(i)}_{NB}$ \\
\specialrule{.1em}{.05em}{.05em}
\lasthline 
 \end{tabular}
\caption{\small{Basis of dimension $d=6$ baryon (and lepton) number conserving operators with a right-handed neutrino $N$ \cite{delAguila:2008ir, Liao:2016qyd}. Here $l_i$, $u _i$, $d _i$ and $L_i$, $Q _i$ denote, for the family labeled $i$, the right handed $SU(2)$ singlet and the left-handed $SU(2)$ doublets, respectively (collectively $f_i$). The field $\phi$ is the scalar doublet, $B_{\mu\nu}$ and $W_{\mu\nu}^I$ represent the $U(1)_{Y}$ and $SU(2)_{L}$ field strengths respectively. Also $\sigma^{\mu \nu}$ is the Dirac tensor, $\gamma^{\mu}$ are the Dirac matrices, and $\epsilon=i\sigma^{2}$ is the anti symmetric symbol in two dimensions. Types S, V and T stand for scalar, vectorial and tensorial (one-loop level generated) structures.}}\label{tab:Operators}
\end{table}

The authors of Ref. \cite{Aparici:2009fh} performed a detailed study of the phenomenology of dimension 5 SMNEFT operators. These include the Weinberg operator $\mathcal{O}_{W}=(\bar{L}\tilde{\phi})(\phi^{\dagger}L^{c})$ \cite{Weinberg:1979sa} which contributes to the light neutrino masses, $\mathcal{O}_{N\phi}=(\bar{N}N^{c})(\phi^{\dagger} \phi)$ which gives Majorana masses and couplings of the heavy neutrinos to the Higgs (its LHC phenomenology has been studied in \cite{Barducci:2020icf, Butterworth:2019iff, Jones-Perez:2019plk, Caputo:2017pit, Graesser:2007yj}), and the operator $\mathcal{O}^{(5)}_{NB}= (\bar{N}\sigma_{\mu \nu}N^{c}) B^{\mu \nu}$ inducing magnetic moments for the heavy neutrinos, which is identically zero if we include just one sterile neutrino $N$ in the theory. As they do not contribute to the processes studied -discarding the heavy-light neutrino mixings- we will only consider the contributions of the dimension 6 operators, following the treatment presented in \cite{delAguila:2008ir, Liao:2016qyd}, and shown in Tab.\ref{tab:Operators}. The effective operators above can be classified by their Dirac-Lorentz structure into \emph{scalar}, \emph{vectorial} and \emph{tensorial}. The couplings of the tensorial operators are naturally suppressed by a loop factor $1/(16\pi^2)$, as they are generated at one-loop level in the UV complete theory \cite{delAguila:2008ir, Arzt:1994gp}.

\subsection{Model implementation}\label{sec:feynrulesimplementation}

In order to estimate the sensitivity reach of the studied processes, we make use of Monte-Carlo (MC) techniques to perform numerical simulations. We build our model in \texttt{FeynRules 2.3}\cite{Alloul:2013bka}, which generates UFO files \cite{Degrande:2011ua} as output, which we then couple to \texttt{MadGraph5\_aMC@NLO 2.5.5} \cite{Alwall:2014hca, Alwall:2011uj}.

We implement the effective interactions in the \texttt{FeynRules} model SMNeff6. As we consider the heavy $N$ to be a Majorana state, the implementation of the model in \texttt{FeynRules} is made compatible with the \texttt{Madgraph5} restrictions on four-fermion operators: as mentioned in its manual \cite{Alwall:2014hca}, \texttt{Madgraph5} can not handle Majorana fermions in operators with more than two fermions. We have circumvented this problem by implementing some particular renormalizable SM extension, which includes the necessary scalar, leptoquark or vector mediator fields, that will generate the operators under consideration in the infrared regime. We are not interested in the auxiliary fields dynamics so we set their masses and decay widths in order to reproduce the behavior we already studied thoroughly for these effective interactions, in particular the decays of the $N$ \cite{Duarte:2016miz, Duarte:2015iba} and its production cross section in 2-2 scattering in different collider environments \cite{Duarte:2018xst, Duarte:2018kiv, Duarte:2016caz}.  

The operators leading to effective lagrangian terms which only involve at most two fermions can be implemented directly in \texttt{FeynRules}. They are those on the first two rows of Tab.\ref{tab:Operators}, which parameterize interactions of the $N$ with the standard vector bosons and the Higgs field and the last row involving tensorial interactions. However, the four-fermion operators in the third and fourth groups in Tab.\ref{tab:Operators} require a specific renormalizable SM extension for \texttt{Madgraph5} to be able to handle them. 

The operator $\mathcal{O}_{duNl}$, leading to the second term in eq.\eqref{eq:lag_tree} is implemented with the mediation of an auxiliary charged vector boson $V_0$ with the lagrangian terms $\bar{N} \gamma^{\mu} P_R  l_i ~V_{0_{i}}$ and $\bar{u}_{j}\gamma^{\mu}P_R d_{j} ~V_{0_{j}}$, which is integrated out numerically by setting its mass to a very high value\footnote{In our numerical implementation all the auxiliary mediators are fixed to have a mass $M=10^{8}~ GeV$ and zero decay width.} and its decay width to zero. As we are interested in the possible family mixings between the lepton and quark sides, we define a $3\times 3$ matrix coupling $\alpha(i, j)_{duNl}$ in flavour space, which allows us to choose between flavour-diagonal or non-diagonal interactions.     

The scalar mediated four-fermion operators $\mathcal{O}_{LNLl}$, $\mathcal{O}_{QuNL}$ and $\mathcal{O}_{LNQd}$ each lead to two lagrangian terms in eq.\eqref{eq:lag_tree}. These can be implemented using neutral ($S_0, ~ S_1, ~S_2$) and charged ($S_{P0}, ~S_{P1}, ~S_{P2}$) scalar auxiliary mediators, which carry a flavour index. For instance, in the case of $\mathcal{O}_{LNLl}$ we include the lagrangian terms $\overline{\nu_{i}}P_R N S_{0_i}$ and $\overline{l_j} P_R l_{j} {S^{\dagger}_{0_j}}$ to generate the first term in the second row of eq.\eqref{eq:lag_tree}, and the terms $\overline{l_i} P_R N {S^{\dagger}_{P0_i}}$ and $\overline{\nu_{i}} P_R l_{i} S_{P0_j}$ to generate the second. We also include for them $3 \times 3$ matrix couplings in flavour space $\alpha(i, j)_{LNLl}$, $\alpha(i, j)_{QuNl}$ and $\alpha(i, j)_{LNQd}$ which let us choose flavour-diagonal or non-diagonal interactions.  

The operator $\mathcal{O}_{QNLd}$ mixes explicitly quarks and leptons in fermion lines. It can be implemented by the mediation of ``leptoquark'' fields we call $S_U$ and $S_D$, with the lagrangian terms $\bar{u} P_R N S_U - \bar{d} P_R N S_D$ and $\bar{l} P_R d \overline{S_U} + \bar{\nu} P_R d \overline{S_D}$. These fields are electrically charged ($Q(S_U)= 2/3$, $Q(S_D)= -1/3$) and carry $SU(3)$ color indices. We make two versions for this operator: choosing to preserve the quark flavour on the four-fermion vertex, or the flavour of the fermion line, the former is the option written in eq.\eqref{eq:lag_tree}. 

The vectorial operators involving two $N$ fields $\mathcal{O}_{fNN}$ are implemented with the mediation of auxiliary massive neutral vector fields $V_f$ (this is done separately for each interaction with a fermion $f= l, ~u, ~d, L,~ Q$). We write lagrangian terms $\frac{1}{2} N \gamma^{\mu} \gamma^{5} N ~V_f$ and $\bar{f} \gamma^{\mu} P_R f ~V_f$, which are coupled to generate four-fermion vertices given by $\mathcal{O}_{fNN}$ in Tab. \ref{tab:Operators}. These operators do not contribute to the processes studied in this work, neither to the $N$ decay, so we do not consider them further. For another implementation of the leptoquark and vector four-fermion interactions see Ref. \cite{Cottin:2021lzz}. See also Ref. \cite{Cirigliano:2021peb} for an implementation of the $\mathcal{O}_{Nl\phi}$ $d=6$ operator together with seesaw mixings at NLO. 

We have checked that our model implementation reproduces our previous results for the $N$ decay width, in every channel \cite{Duarte:2015iba, Duarte:2016miz}, and also the cross sections for every 2-2 four-fermion scattering involving one $N$ field. 

\subsection{Constraints on effective couplings}\label{sec:constraints}

Recent works on right-handed effective neutrino interactions (SMNEFT), including dimension 6 operators have derived constraints on the different effective couplings values $\alpha_{\mathcal{J}}$, or alternatively on the new physics scale $\Lambda$. Most constraints are valid for Majorana $m_N$ masses below the benchmark points we consider in this work: $m_N = 50, 100, 150$ GeV.
 
Reference \cite{Alcaide:2019pnf} sets bounds on four-fermion effective couplings with one and two $N$ fields, obtaining values below unity for $\Lambda=1$ TeV and masses $m_N \lesssim 1$ GeV. Reference \cite{Biekotter:2020tbd} studies the SMNEFT includying also Type I seesaw mixings for the $N$ with SM light neutrinos. They study the mass regime in which the decay channel $N\to \nu \gamma$ is dominant, for $m_N \lesssim 10$ GeV. More recently, in Ref. \cite{Beltran:2021hpq}, the authors obtain exclusion limits for the sensitivity reach in the $m_N- \Lambda$ plane discarding the heavy-light mixings and using effective operators with one $N$, also for masses $m_N$ below $50$ GeV, proposing a displaced-vertex search strategy for the high luminosity LHC. Also, the authors in Ref. \cite{Cottin:2021lzz} study operators with two $N$ fields and Type I seesaw mixings, showing sensitivity reaches in the $m_N-\Lambda$ plane for masses up to $1$ TeV for fixed values of the heavy-light mixings (which drive the $N$ decay for the considered scenarios), using displaced $N$ decays in the LHC with ATLAS and several far detector experiments. As these operators do not contribute to the processes studied in this work, we do not restrict our effective couplings to these constraints. Our group has obtained bounds for the effective couplings, separating vector, scalar and tensor contributions, considering $N$-mediated B meson decays in LHCb and Belle, which also apply for $m_N< 5$ GeV \cite{Duarte:2019rzs}.

In Ref. \cite{Cirigliano:2021peb} the authors obtain a bound on the leptonic right handed current $\mathcal{O}^{(2)}_{LN\phi}$ coupling for $m_N=100 ~(1000)$ GeV from CMS results \cite{CMS:2018iaf}, assuming the $N$ only decays via seesaw mixings. It can be translated, when considering only one massive heavy neutrino state (almost coinciding with the sterile neutrino state), into the bound $\alpha^{(2)}_W \lesssim 2 ~(20)$, for $\Lambda=1$ TeV, considering the $\mu N W$ interaction. The bounds for the first fermion family ($\alpha^{(1)}_W$) are $1.5-2$ times weaker. These bounds are well above the numerical values considered in our analysis.  
    
For this work we do take into account existing constraints \cite{KamLAND-Zen:2016pfg} on the operators contributing to neutrinoless double beta decay ($0\nu \beta \beta$-decay) following the treatment in Ref. \cite{Duarte:2016miz} and we set the bound $\alpha_{0\nu\beta\beta}=3.2\times 10^{-2}\left(\frac{m_N}{100 GeV} \right)^{1/2}$ for $\Lambda=1$ TeV, for the couplings of the $\mathcal{O}^{(1)}_{LN\phi} , ~\mathcal{O}^{(1, 1)}_{duNl}, ~\mathcal{O}^{(1, 1)}_{QuNL}, ~\mathcal{O}^{(1, 1)}_{LNQd} , ~\mathcal{O}^{(1, 1)}_{QNLd} $ operators. These appear as contributions to the $\Gamma_N$ total width, and to the $N$ production vertex in the considered processes. We do not impose constraints on the other effective operators.

\section{Collider analysis}\label{sec:Collider analysis}

\begin{figure*}[h]
 \centering
  \subfloat[]{\label{fig:ee_nuN}\includegraphics[totalheight=5.8cm]{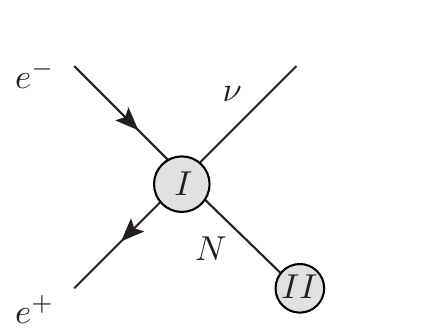}}~
  \subfloat[]{\label{fig:n_lepysemilep}\includegraphics[totalheight=5.8cm]{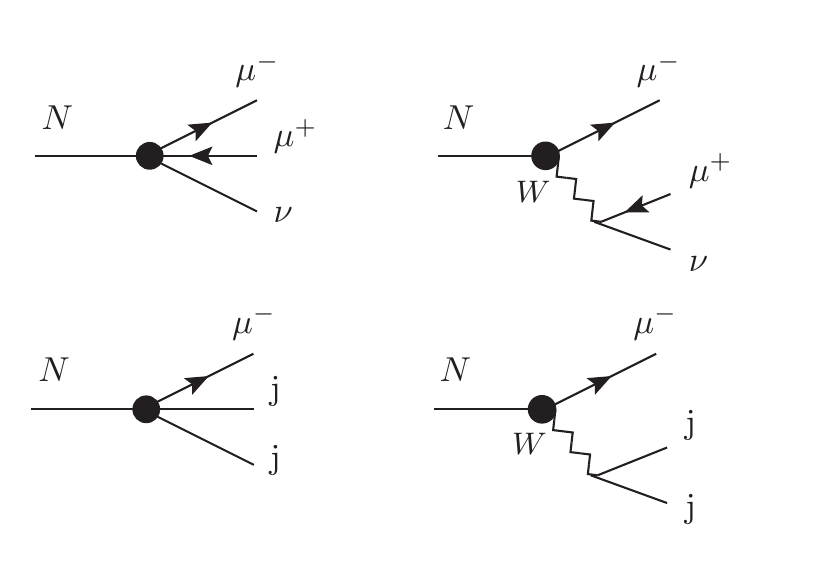}}
 \caption{ \label{fig:ee_muons}{Single $N$ production and decay in the di-muon channel and di-jet channels.} }
 \end{figure*}

We consider the single production of Majorana neutrinos $N$ in the process $e^{+}e^{-} \to \nu N$, followed by a purely leptonic decay $N \to \mu^{-} \mu^{+} \nu$ or a semi-leptonic decay $N \to \mu^{-} \mathrm{j} \mathrm{j} $. The single $N$ production mode in an $e^{+} e^{-}$ collider in our simplified effective interactions model, depicted in Fig.\ref{fig:ee_nuN}, is dominated by the $\mathcal{O}^{(1,1)}_{LNLl}$ operator contribution, because of the strong constraint imposed on the alternative $\mathcal{O}^{(1)}_{Nl\phi}$ interaction by the non-observation of the neutrinoless double beta decay, as discussed in Sec.\ref{sec:feynrulesimplementation}.

We start by studying the pure leptonic process $e^{+}e^{-} \to \nu \mu^{-} \mu^{+} \nu $, where the Majorana $N$ is produced in the primary vertex together with a light neutrino, and subsequently decays into a di-muon pair and a light neutrino, as depicted in the upper diagrams in Fig.\ref{fig:n_lepysemilep}. The decay of the $N$ in vertex $(II)$ to a di-muon pair and a light neutrino involves the contributions of the operators $\mathcal{O}^{(2,2)}_{LNLl}$ and $\mathcal{O}^{(2)}_{Nl\phi}$, now for the second family flavour.    

Also, we aim to study the semi-leptonic process $e^{+}e^{-} \to \nu \mu^{-} \mathrm{j} \mathrm{j}$, where the Majorana $N$ decays into a muon and two jets. While the $N$ production mechanism remains unchanged, the decay of the $N$ in vertex $(II)$ now involves the contributions of the vector four-fermion operator $\mathcal{O}^{(2,i)}_{duNl}$ together with the vector $\mathcal{O}^{(2)}_{Nl\phi}$ and the scalars $\mathcal{O}^{(2, i)}_{LNLl}$, $\mathcal{O}^{(2, i)}_{QuNL}$, $\mathcal{O}^{(2, i)}_{LNQd}$ and $\mathcal{O}^{(i, 2)}_{QNLd}$, as shown in the lower diagrams in Fig.\ref{fig:n_lepysemilep}. 

We can thus explicitly take into account the following tree-level generated effective lagrangian terms involved in those processes:
\begin{eqnarray}\label{eq:lag_tree}
  \mathcal{L}&&= \frac{1}{\Lambda^2} \sum_{i,j}\Big\{ -\alpha^{(i)}_W \frac{ ~v ~m_W}{\sqrt{2}}\, \overline{l_i} \gamma^{\nu} P_R N  \, W^{-}_{\mu}   		  
	+	\alpha^{(i, j)}_{V_0} \, \overline{u _j} \gamma^{\nu} P_R d _j \,  \, \overline{l_i} \gamma_{\nu} P_R N  
		 \nonumber \\
		&& 
		 + \alpha^{(i, j)}_{S_0}(\bar \nu_{i} P_R N  \,  \,  \overline{ l_{j}} P_R l_{j}-  \overline{ \nu_{j}} P_R l_{j}    \,  \, \overline{l_{i}} P_R N) 
		  	+	\alpha^{(i, j)}_{S_1}\, (\overline{u _j} P_L d _j \, \, \overline{l_i} P_R N + \overline{u _j} P_L u _j \, \, \overline{\nu_i} P_R N) 
		\nonumber \\		
		&&  +  \alpha^{(i, j)}_{S_2}\, ( \overline{d _j} P_R d _j \, \, \overline{\nu_i} P_R N- \overline{u _j} P_R d _j \, \, \overline{l_i} P_R N ) 
		+ \alpha^{(i,j)}_{S_3}\, (\overline{u _i} P_R N \, \,  \overline{l_j} P_R d _i - \overline{d _i} P_R N \, \,  \overline{\nu_j} P_R d _i)   
\nonumber \\
		&&
		+ \mbox{h.c.} \Big\}.
\end{eqnarray}
The tree-level $N$ production cross section (Fig.\ref{fig:ee_nuN}) can be calculated as 
\begin{equation}
\sigma(e^+e^-\to \nu_i N)= \frac{(m_N^2-s)^2}{8 \pi \Lambda^4 s} \left\{{\alpha_{S_0}^{(i,1)}}^{2} \frac{(m_N^2+2s)}{8 s} + {\alpha_{W}^{(1)}}^2 \frac{m_{W}^2}{(s +m_W^2 - m_N^2)}\right\},
\end{equation}
while the pure-leptonic and semi-leptonic tree-level $N$-decay widths can be found in Appendix B in \cite{Duarte:2016miz}. 

The irreducible SM backgrounds for both the pure leptonic $e^{+}e^{-} \to \nu \mu^{-} \mu^{+} \nu $ and the semi-leptonic $e^{+}e^{-} \to \nu \mu^{-} \mathrm{j} \mathrm{j}$ processes involve diagrams with intermediate standard vector bosons (photons, $Z$) and Higgs bosons in $s$ channels, which subsequently decay into muon pairs, light neutrino pairs, or quark pairs, and $W$ bosons decaying leptonically or hadronically. The dominant SM backgrounds for both processes are events that come from $e^{+}e^{-} \to W^{-}W^{+}$, with both $W$ decaying leptonically in the first case \cite{Hernandez:2018cgc, Banerjee:2015gca}, and the $W^{+}$ decaying hadronically in the second \cite{Liao:2017jiz}.

As we want to tackle the single $N$ production, which brings at least one undetectable light neutrino in the final state, we cannot rely on explicitly observing lepton number violation as evidence for the Majorana $N$ production. Given that our signals include the $N \to \mu^{-} W^{+}$ channel coming from the $\mathcal{O}^{(2)}_{Nl\phi}$ contribution, we cannot just rely either on the invariant mass distribution of the jet pair, or the transverse mass of the $\nu \mu^{+}$ pair (which peak at the $W$ mass for the dominant SM background) to impose cuts \cite{Banerjee:2015gca, Antusch:2016ejd, Liao:2017jiz}. Alternatively, we will rely on the signal topology and kinematics, where depending on its energy and mass the boosted $N$ will produce its daughter particles in a collimated cone, and exploit the very good angular and energy resolution at future electron-positron colliders to construct forward-backward asymmetries between the leptons (and jet) in the final states to disentangle the $N$ contribution.

The proposed signals can be studied in future lepton colliders like the linear ILC \cite{Behnke:2013xla} or circular colliders like the FCC-ee \cite{dEnterria:2016sca} and the CEPC \cite{CEPCStudyGroup:2018rmc}. For concreteness, throughout the paper we will consider an $e^{+}e^{-}$ collider with center of mass energy $\sqrt{s}=500 $~GeV and integrated luminosity $\mathcal{L}=500 ~fb^{-1}$ for estimating the numbers of events. These values correspond to one of the proposed ILC operation modes \cite{Baer:2013cma}.

In order to study the prospects of measuring the single $N$ production with pure leptonic and semi-leptonic decays we have implemented the effective interactions model in \texttt{FeynRules 2.3} \cite{Alloul:2013bka}, as described in Sec.\ref{sec:feynrulesimplementation}. The UFO output \cite{Degrande:2011ua} was input into \texttt{MadGraph5\_aMC@NLO 2.5.5}  \cite{Alwall:2014hca, Alwall:2011uj}, and we generate LHE events at parton level, which are read by the embedded version of \texttt{PYTHIA 8}  \cite{Sjostrand:2014zea}, and then are interphased to \texttt{Delphes 3.5.0} \cite{deFavereau:2013fsa} with the DSiDi card \cite{Potter:2016pgp} for a fast detector simulation. The analysis of the generated events at the reconstructed level is made with the expert mode in \texttt{MadAnalysis5 1.8.58} \cite{Conte:2012fm}. 

For concreteness, we explore a few benchmark scenarios, and simplify the parameter space setting all the effective couplings $\alpha_{\mathcal{J}}$ in eq. \eqref{eq:lagrangian} and Tab.\ref{tab:Operators} to the same numerical value $\alpha$ (except for the operators with charged leptons of the first family constrained by the $0\nu\beta\beta$-decay bound, as explained in sec. \ref{sec:feynrulesimplementation}). We also fix the new physics scale $\Lambda=1$ TeV and show our results for different $N$ mass values $m_N$. 

We generate events by asking for the processes $e^{+}e^{-} \to \nu \mu^{-} \mu^{+} \nu $ (pure-leptonic) and $e^{+}e^{-} \to \nu \mu^{-}  \mathrm{j}  \mathrm{j}$ (semi-leptonic), with $\nu=\nu_{\ell}, \bar{\nu_{\ell}}, ~\ell=e,\mu, \tau$, and only light quark jets ($\mathrm{j}= u, d, c, s$). These events include the signals with the production of an intermediate $N$ state (even in $t$-channels), as well as the pure-SM processes, allowing for interference (signal plus background events: S+B). We also generate signal-only events (S), by producing the heavy $N$ explicitly by $e^{+}e^{-} \to \nu N$ and asking for the leptonic $N\to \mu^{-}\mu^{+} \nu$ or semi-leptonic $N \to \mu^{-} \mathrm{j} \mathrm{j}$ decays. 
\begin{figure*}[t]
\centering
\subfloat[]{\label{fig:xs_mus}\includegraphics[width=0.5\textwidth]{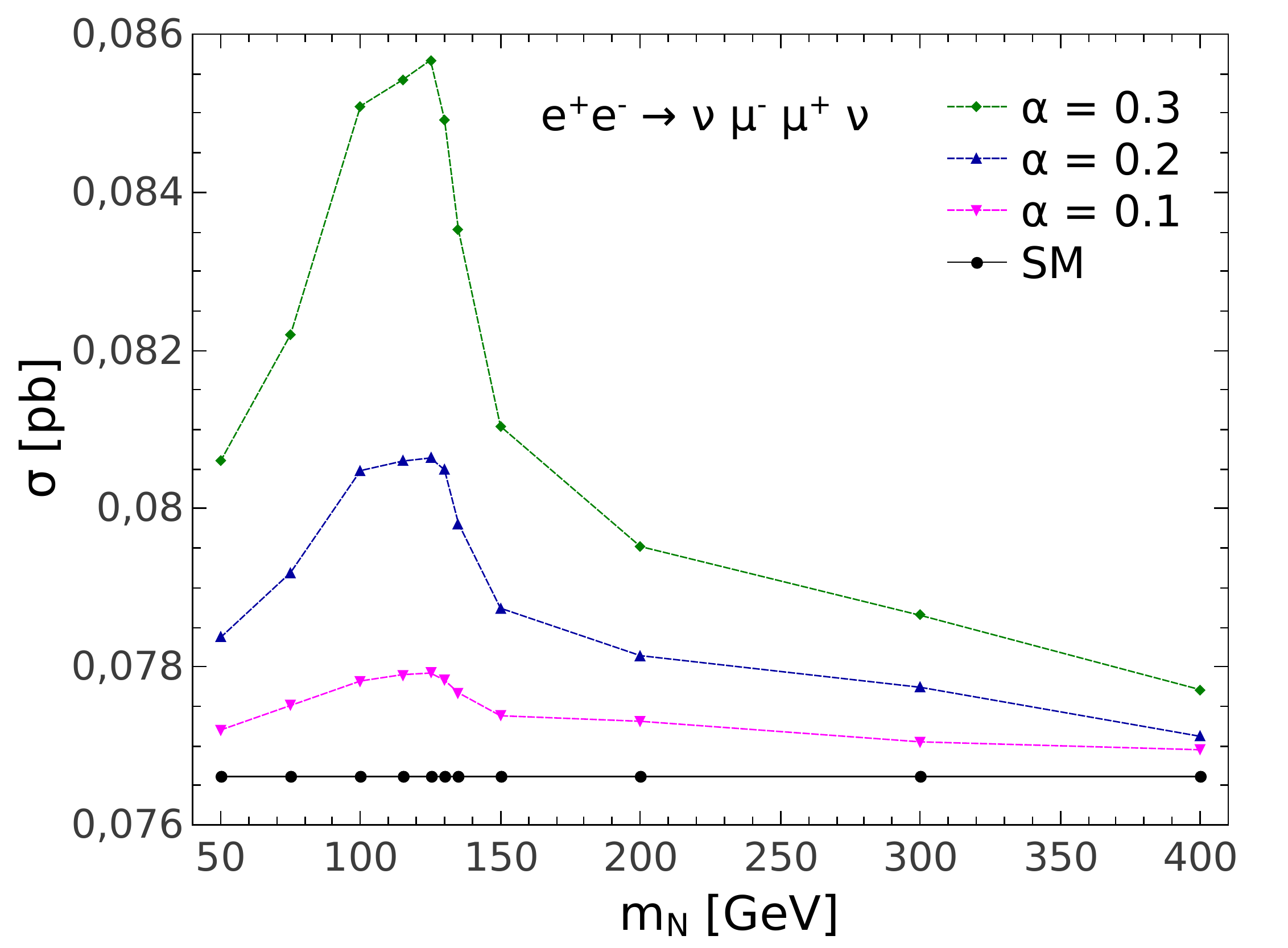}}~
\subfloat[]{\label{fig:xs_jets}\includegraphics[width=0.5\textwidth]{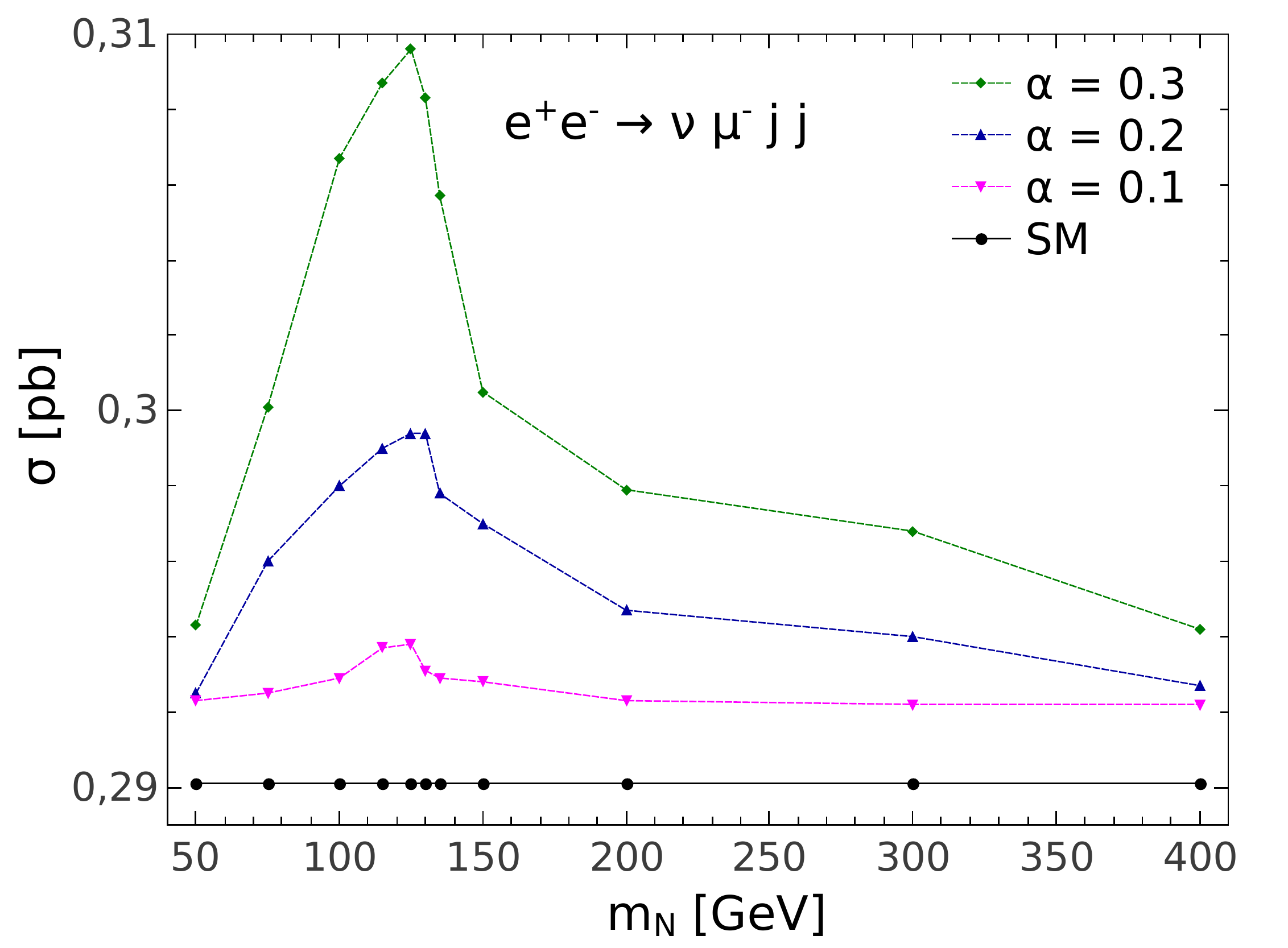}}
 \caption{ \label{fig:crossect}{Parton-level cross sections for the pure leptonic (a) and semi-leptonic (b) processes, with acceptance cuts (see text).} }
 \end{figure*}

We adopt the following basic acceptance cuts for both the pure-leptonic and the semi-leptonic process: we keep transverse momenta for jets $p_{T}^{ \mathrm{j}} > 20$ GeV and leptons $p_{T}^{\ell} > 10 $ GeV, pseudorapidities $|\eta_{\mathrm{j}}|<5$, $|\eta_{\ell}|< 2.5$, and isolation between jets and leptons $\Delta R_{\mathrm{j} \mathrm{j}, \ell \ell, l \mathrm{j}} > 0.4$. 

The number of events and cutflows for the signal-only (S), signal plus background (S+B, generated with interference) and SM background-only (B) datasets are shown in Tab.\ref{tab:cutflow_muons} for the pure-leptonic and Tab.\ref{tab:cutflow_jets} for the semi-leptonic processes.   

In Fig.\ref{fig:crossect} we show the parton-level cross section values we obtain for both processes, generating signal plus background (S+B) events datasets with different values of the effective couplings $\alpha$, together with the SM background-only (B) values
\footnote{We have checked that our SM value for the cross section $\sigma_{SM}(e^{+}e^{-} \to \nu \mu^{-}  \mathrm{j}  \mathrm{j})$ agrees with \cite{Liao:2017jiz} (using their cuts and the CEPC Delphes card).}. 
For $m_N$ near $120$ GeV, we find an enhancement of the signal cross section, due to the contribution of the $\mathcal{O}^{(i)}_{Nl\phi}$ operator to the $N$ production and decay when the $W$ is on-shell.

\subsection{Leptonic channel}\label{sec:Leptonic}

For the pure leptonic mode $e^{+}e^{-} \to \nu \mu^{-} \mu^{+} \nu $, with two light neutrinos in the final state, we exploit the precision in measurements of the muons pair momenta. As the $N$ is produced together with a light neutrino in a 2-2 process, the energy and boost of the $N$ are completely determined in the CM frame for each mass value $m_N$. Its production will be reflected in the dependence of the various observables on the summed energy of the di-muon pair $E_{\mu \mu}=E(\mu^{-}+\mu^{+})$. 

\begin{figure*}[t]
\centering
\subfloat[]{\label{fig:DR_muons_ss}\includegraphics[width=0.5\textwidth]{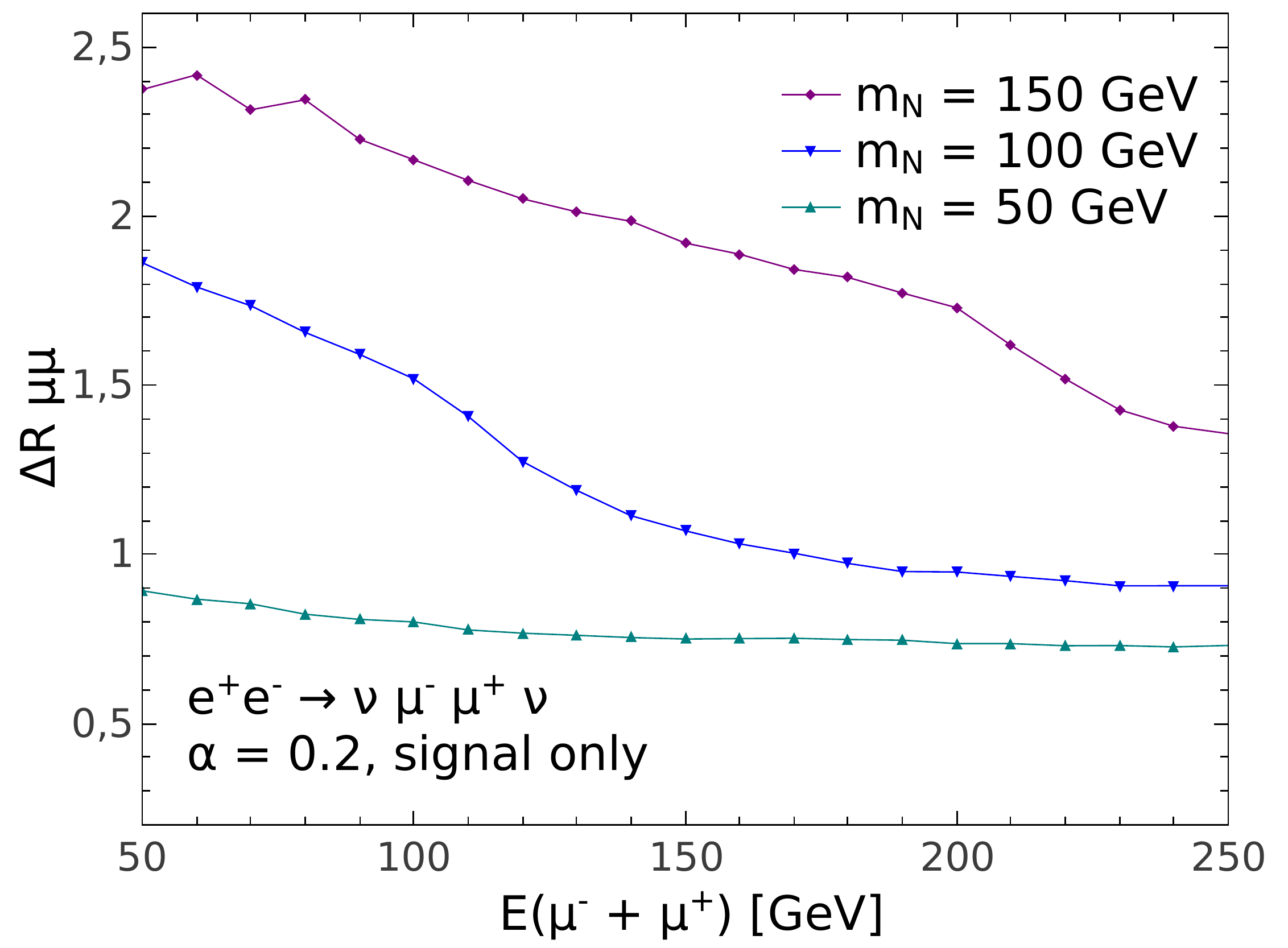}}~
\subfloat[]{\label{fig:AFB_muons_ss}\includegraphics[width=0.5\textwidth]{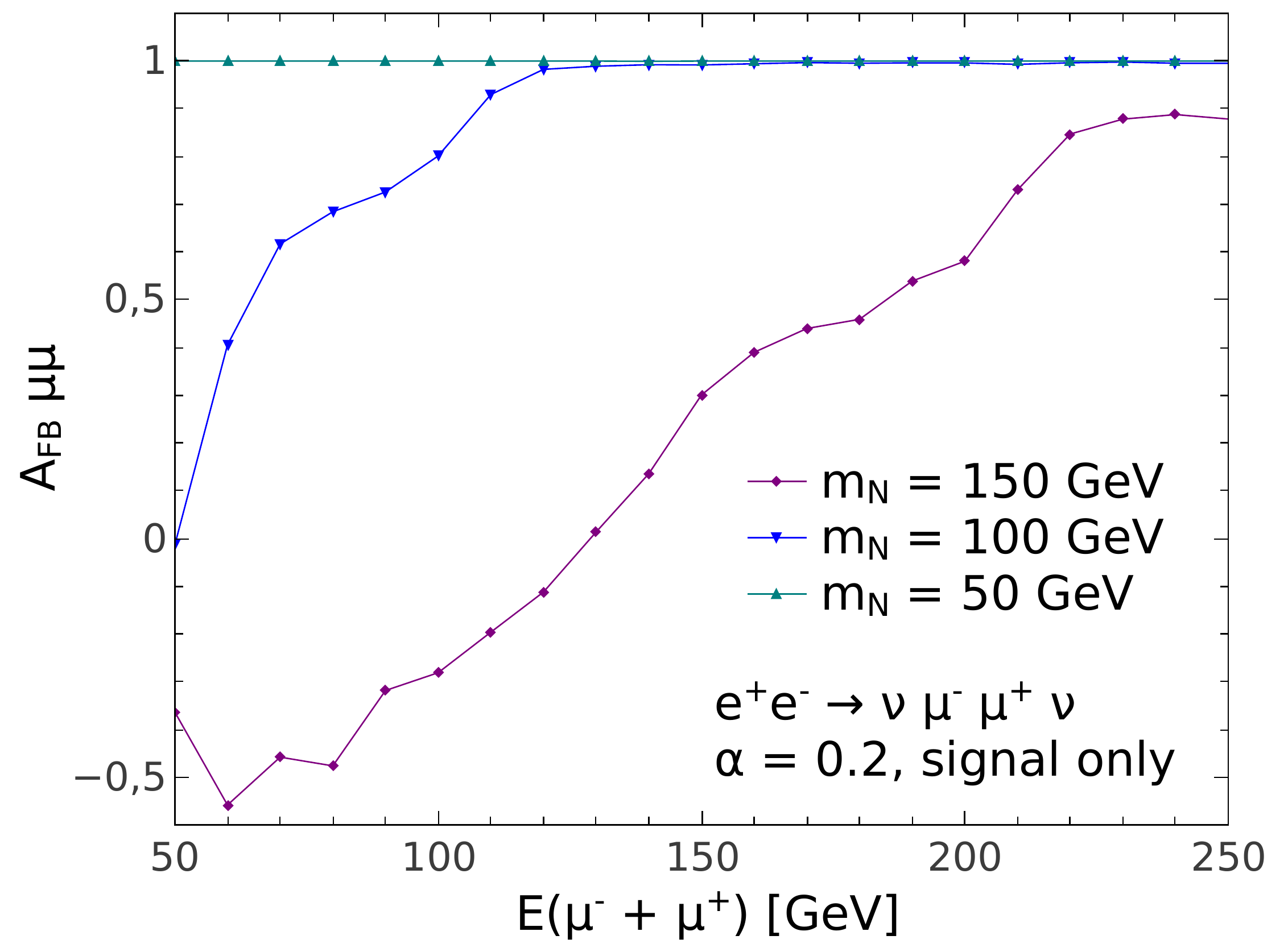}}
 \caption{ \label{fig:DR_AFB_muons_SS}{Average $\Delta R_{\mu \mu}$ (a) and $A^{FB}_{\mu \mu}$ (b) distributions with $E_{\mu \mu}$, signal-only (S) no cuts.} }
 \end{figure*}

A preliminary test to see the dependence of the di-muon pair separation in $ \Delta R = \sqrt{\Delta \eta^{2}+\Delta \phi^{2}}$ with $E_{\mu \mu}=E(\mu^{-}+\mu^{+})$ and the intermediate $N$ mass $m_N$ is shown in Fig.\ref{fig:DR_muons_ss}. We calculate the $\Delta R_{\mu \mu}$ value averaged over events for each $E_{\mu \mu}$ bin. Here we generate signal-only (S) events producing the $N$ and setting its decay $N \to \mu^{+} \mu^{-} \nu$ explicitly, for some benchmark mass values and couplings $\alpha=0.2$. 

As expected, we find that for low $m_N = 50$ GeV the muons come out roughly in the same direction, and with growing $m_N$ they are less boosted and start to separate. Also, for a fixed value of $m_N$, with increasing energy, the muons are more boosted, and $\Delta R_{\mu \mu}$ diminishes. 

In order to analyse this behavior more precisely, we consider a forward-backward asymmetry $A_{ \mu \mu}^{FB}$ between the flight directions of the outgoing muon and anti-muon, for different values of their summed energies $E_{\mu \mu}$. We thus define: 
\begin{equation}\label{eq:Amunu}
 A_{ \mu \mu}^{FB} = \frac{N_{+}-N_{-}}{N_{+}+N_{-}}, 
\end{equation}
where $N_{\pm}$ is the number of events with a positive (negative) value of $\cos(\theta)$, the angle between the final muon and anti-muon flight directions in the Lab (or CM) frame in the $e^{+}e^{-}$ collision. We plot $A_{ \mu \mu}^{FB}(E_{\mu \mu})$ for signal-only (S) events in Fig.\ref{fig:AFB_muons_ss}, for the same benchmark mass values and couplings, confirming the aforementioned behavior: for low $N$ mass the muons are boosted and emerge in the same direction ($A_{ \mu \mu}^{FB}=1$) and tend to separate ($A_{ \mu \mu}^{FB}<1$) even coming out mostly in opposite directions ($A_{ \mu \mu}^{FB}<0$) when $m_N$ increases for low values of the summed energies of the outgoing muons. When their energy increases, the muon pairs tend to be very boosted, giving a positive asymmetry value. 

\begin{figure*}[h]
\centering
\subfloat[]{\label{fig:E_muons_sb}\includegraphics[width=0.5\textwidth]{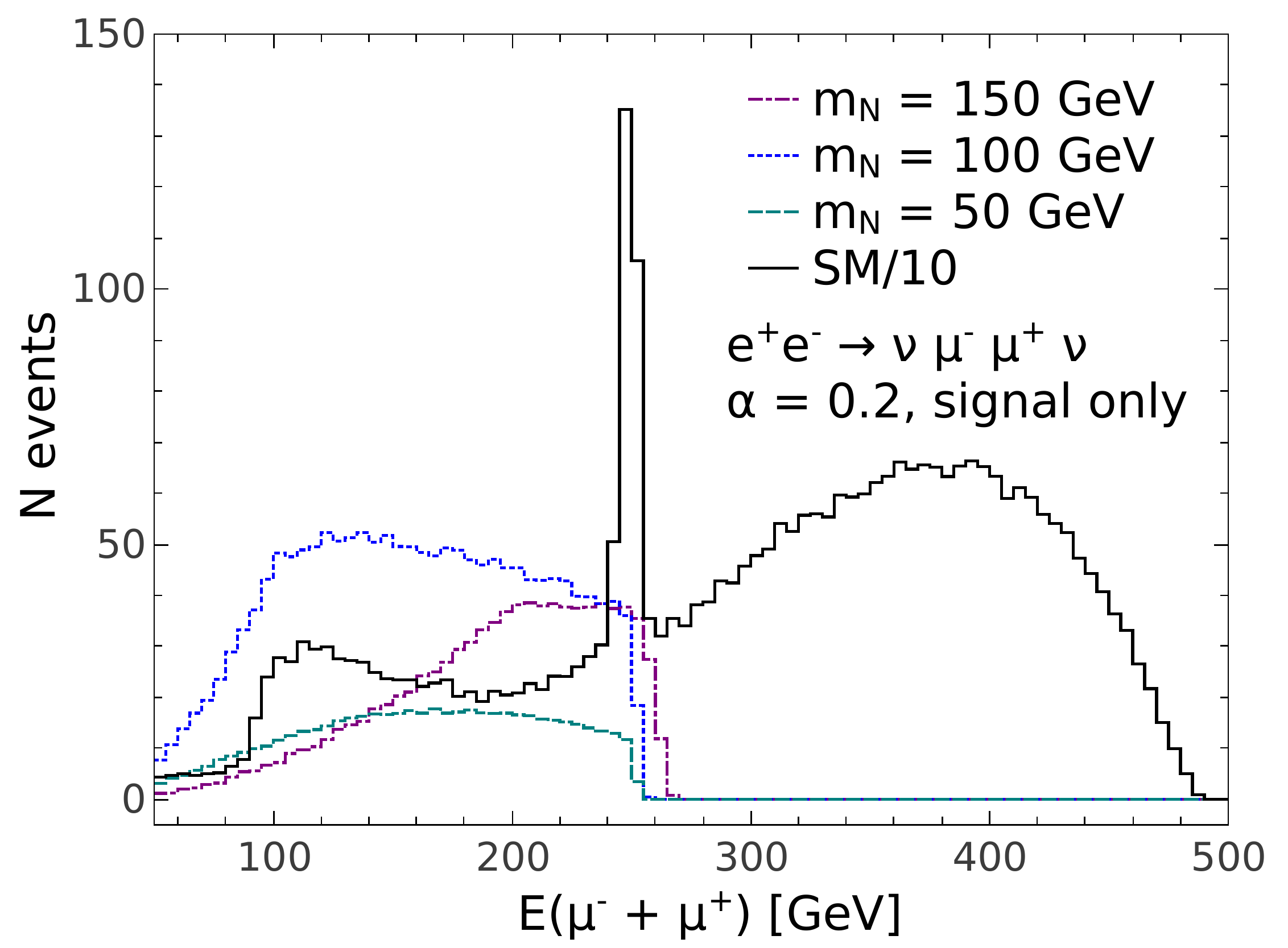}}~
\subfloat[]{\label{fig:afb_muons}\includegraphics[width=0.5\textwidth]{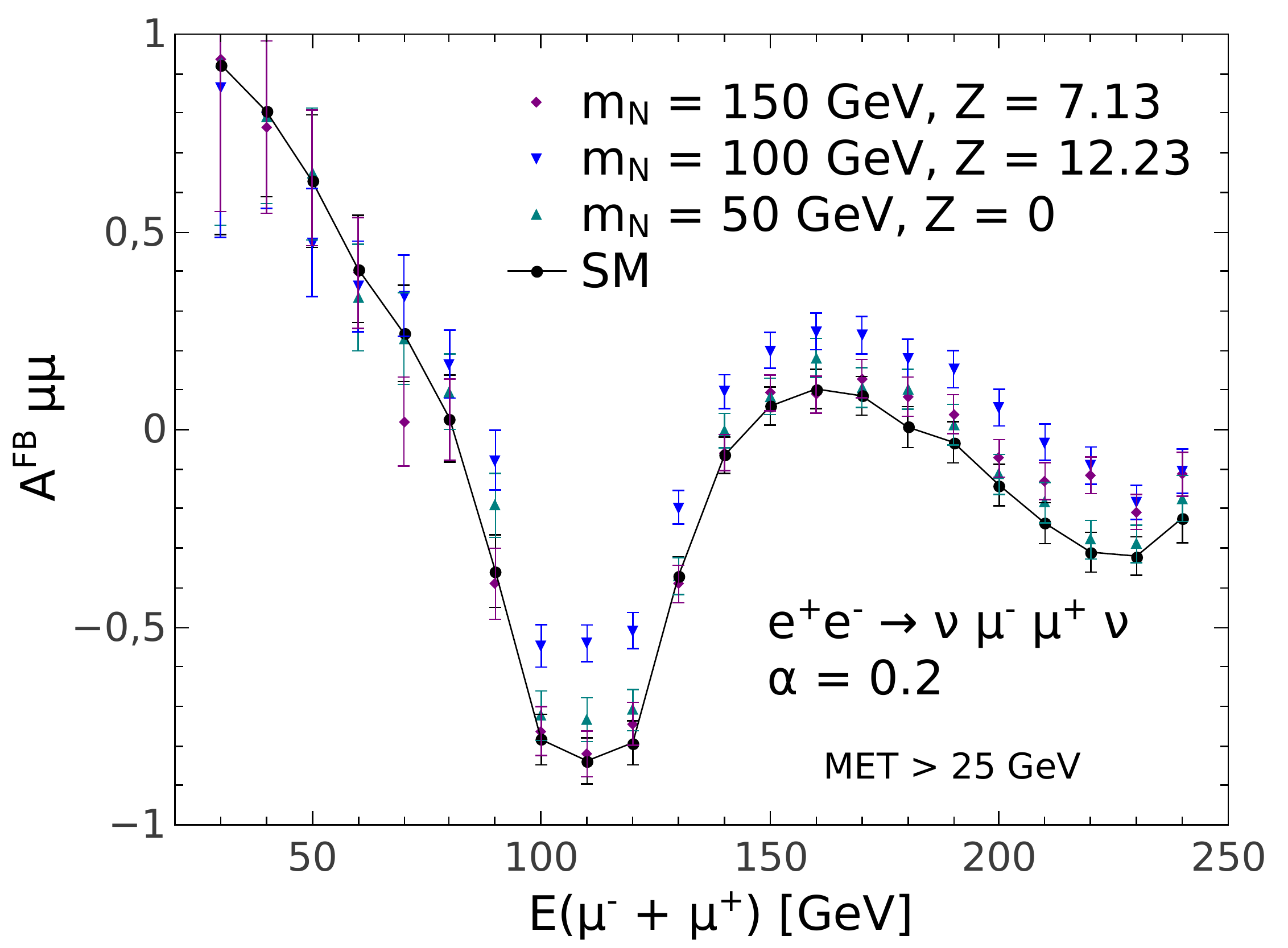}}
\caption{ \label{fig:E_AFB_muons}{Number of events (a) and $A^{FB}_{\mu \mu}$ (b) distributions with $E_{\mu \mu}$, for $\alpha=0.2$, $\Lambda=1$TeV, $\sqrt{s}=500$ GeV, $\mathcal{L}=500 ~fb^{-1}$. }}
\end{figure*}

With these hints on the signals-only kinematical behavior, we explore the possibility to use this forward-backward asymmetry to separate the events in the realistic dataset generated with the SMNeff6 UFO and allowing for effective-SM interference (S+B) from the SM-only (B) events for the pure leptonic process $e^{+}e^{-} \to \nu \mu^{-} \mu^{+} \nu $. We study different kinematical variables distributions for the process in order to separate the signals from the SM background. We apply a selection cut keeping events with a minimum value of missing transverse energy (MET > $25$ GeV), accounting for neutrinos in the final state. The distribution of the signal-only (S) and background-only (B) number of events with $E_{\mu \mu}$ is shown in Fig.\ref{fig:E_muons_sb}. We find that a cut on $E_{\mu \mu}< 240$ GeV helps to separate the signals from the dominant SM background, which peaks at $E_{\mu \mu}= \sqrt{s}/2$, in a symmetric configuration where the muon anti-muon pair shares half the energy with the unobservable light neutrinos. The corresponding number of events cutflow is shown in Tab.\ref{tab:cutflow_muons}. 
{\footnotesize{
\begin{table}[t]
 \centering
 \begin{tabular}{l  l l  l l  l l  l}
 \firsthline 
 \specialrule{.1em}{.05em}{.05em}
$\alpha= 0.2$  & \multicolumn{2}{l}{$m_N=50$ GeV} & \multicolumn{2}{l}{$m_N=100$ GeV} & \multicolumn{2}{l}{$m_N=150$ GeV}  & SM \\
Cuts:  &  S & S+B &  S  & S+B  &  S & S+B & B \\ 
\specialrule{.03em}{.03em}{.03em}
Muons		& $544$ 	&  $33545$  & $1659$ & $34612$ & $901$ & $33860$ & $33060$ \\  
\specialrule{.02em}{.02em}{.02em}  
 MET  &  $539$(99\%) &   $27098$(81\%) & $1617$(97\%) & $28035$(81\%) & $884$(98\%)&  $27296$(81\%)&  $26543$(80\%) \\
\specialrule{.02em}{.02em}{.02em}
  $E_{\mu \mu}$ &  $511$(94\%) &  $7085$(21\%) & $1524$(92\%) & $8070$(23\%) & $734$(81\%)  & $7351$(22\%) & $6771$(20\%) \\
 \specialrule{.03em}{.03em}{.03em}
  $S/\sqrt{(S+B)}~~$  &  $6.1$  &    & $16.9$ &   & $8.6$ &  &  \\
\specialrule{.1em}{.05em}{.05em}
\lasthline 
 \end{tabular}
\caption{\small{$e^{+}e^{-} \to \nu \mu^{-} \mu^{+} \nu $ (pure-leptonic). Number of events for signal-only (S), signal and SM background (S+B, generated with interference) and SM background (B), for $\sqrt{s}=500$ GeV, $\mathcal{L}$=500$fb^{-1}$. Muons cut: one muon and anti-muon in final state. MET cut: select events with MET > 25 GeV. $E_{\mu \mu}$ cut: select events with $E_{\mu \mu}$<240 GeV (see Fig.\ref{fig:E_muons_sb}). Signal significance: $S/\sqrt{(S+B)}$ for event-counting experiment.}}\label{tab:cutflow_muons}
\end{table}
}}

Although the signal-only (S) events seem to be able to be discovered with a sensitivity greater than $5\sigma$ in an event-counting experiment with effective couplings $\alpha=0.2$ applying the MET an $E_{\mu \mu}$ cuts, we want to test the possibility to discover the signals in the full (S+B) dataset. So we take one step forward, and use the asymmetry in eq.\eqref{eq:Amunu} to perform a $\Delta \chi^2$ test to find the statistical significance of the separation between the signal plus background ($A_{\mu \mu}^{S+B}$) and background-only ($A_{\mu \mu}^{B}$) data sets for the asymmetry values. 

We build a $\Delta \chi^2$ function as 
\begin{equation}
 \Delta \chi^2= \sum_{E_i} \frac{ \left(A_{\mu\mu}^{S+B}(E_i)- A_{\mu \mu}^{B}(E_i, m_N,\alpha)\right)^2} {\left(({\Delta A_{\mu\mu}^{S+B}})^2 + ({\Delta A_{\mu\mu}^{B}})^2\right)},
\end{equation}
where $E_{i} = E_{\mu \mu}$, the muons summed energy bins shown in Fig.\ref{fig:DR_AFB_muons_SS} and Fig.\ref{fig:E_AFB_muons}. We sum in quadrature the error on the number of events for both asymmetry values in each summed energy bin, $({\Delta A_{\mu \mu}^{B}})^2 + ({\Delta A_{\mu\mu}^{S+B}})^2$, which are taken to be Poisson distributed \cite{Duarte:2018xst}. 

In Fig.\ref{fig:afb_muons} we show the values and error bars of the asymmetry $A_{\mu \mu}^{S+B}$ and $A_{\mu \mu}^{B}$ in $E_{\mu \mu}$ bins, for distinct $m_N$ values and for $\alpha=0.2$. The label shows the value $Z$ of the number of standard deviations $Z \sigma$ of statistical significance we obtain for each dataset with the $\Delta \chi^2$ test. The signal plus background (S+B) events dataset can be well separated from the SM background-only events (B) when $m_N=100$ giving $12 \sigma$, and $m_N= 150$ GeV giving $7 \sigma$, but not for $m_N=50$ GeV, due to the lower cross section of this signal, as can be seen in Fig.\ref{fig:xs_mus}. We find this lower mass value could be resolved with $Z=8$ for couplings $\alpha=0.25$ using the same MET and $E_{\mu \mu}$ cuts.

\subsection{Semi-leptonic channel}\label{sec:Jets}

For the semi-leptonic process $ e^{+}e^{-} \to \nu \mu^{-}  \mathrm{j}  \mathrm{j}$ we aim to see the imprints of the intermediate $N$ on its decay products by measuring the asymmetry $A_{ \mathrm{j} \mu}^{FB}$ between the flight directions of the final muon and the jet with the highest transverse momentum, defined equivalently to eq.\eqref{eq:Amunu}. We expect to see the dependence of the asymmetry with the energy of the muon $E_{\mu}$ for distinct $m_N$ and $\alpha$ values. 

\begin{figure*}[h]
\centering
\subfloat[]{\label{fig:DeltaR_jets_SS}\includegraphics[width=0.5\textwidth]{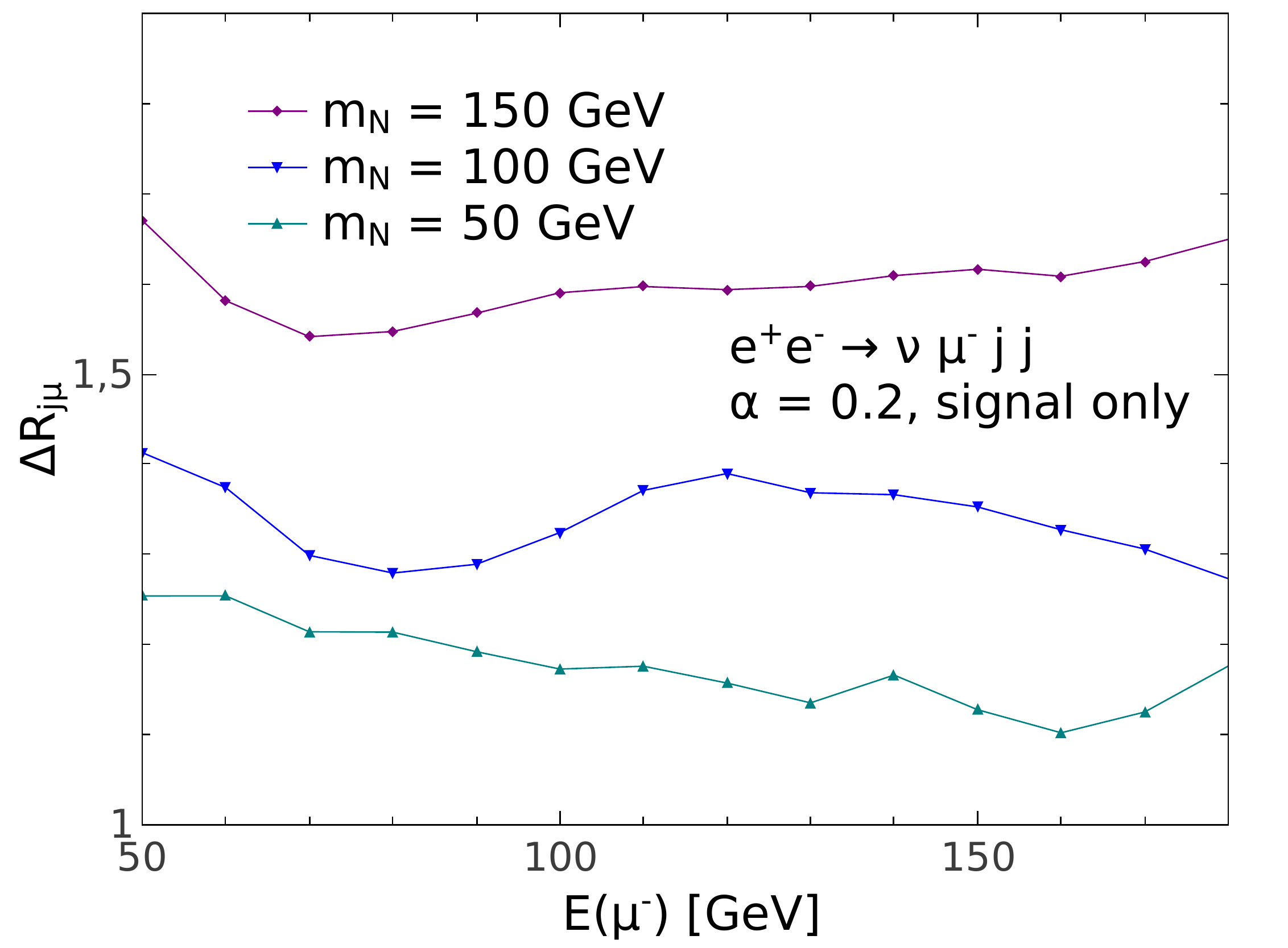}}~
\subfloat[]{\label{fig:AFB_jets_SS}\includegraphics[width=0.5\textwidth]{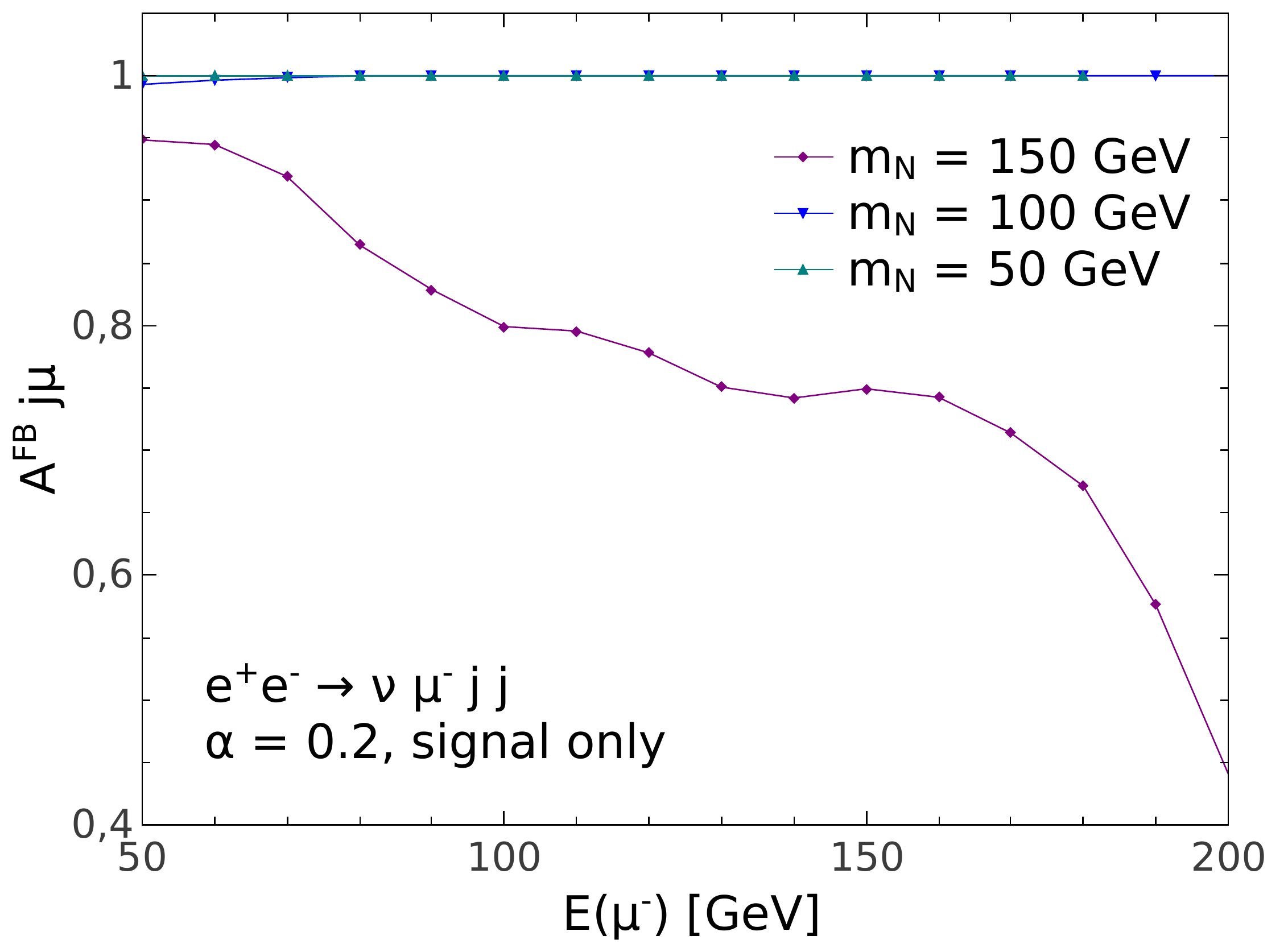}}
\caption{ \label{fig:DR_AFB_jets_SS}{Average $\Delta R_{\mathrm{j} \mu}$ (a) and $A^{FB}_{\mathrm{j} \mu}$ (b) distributions with $E_{\mu}$, signal-only (S) no cuts.}}
\end{figure*}

As we did with the pure leptonic channel, in Fig.\ref{fig:DeltaR_jets_SS} we plot the distribution of the average value of the separation $\Delta R_{\mathrm{j} \mu}$ between the muon and the most energetic jet in each muon energy bin $E_{\mu}$ for signal only events, before applying any cuts. We find that the muon and jet separation depends mostly on the mass of the intermediate Majorana neutrino $m_N$: for low mass the products of the decay $N \to \mu^{-} \mathrm{j} \mathrm{j}$ are more boosted and emerge in a narrower cone. This effect increases slightly with higher muon energy and, in fact, we do not have events in the higher $E_{\mu}$ bins for the $m_N=50$ GeV signal. In the $m_N=150$ GeV signal events, the muon and the jet are more separated. In Fig.\ref{fig:AFB_jets_SS} we plot the asymmetry $A_{ \mathrm{j} \mu}^{FB}$ for signal-only events before cuts. Consistently, the events with a lighter $N$ produce a maximal value $A_{ \mathrm{j} \mu}^{FB}=1$, showing that the muon and jet emerge in the same direction. This changes for the higher $m_N$ values. Here the $N$ is less boosted and the muon and jet emerge in an increasingly open angle for higher muon energies. This can be seen considering that the more energetic muons tend to emerge in the direction of the $N$, and the jets go in the opposite direction (they mostly come from an on-shell $W$, see the right-bottom diagram of Fig.\ref{fig:n_lepysemilep}).    

\begin{figure*}[h]
\centering
\subfloat[]{\label{fig:MTMET_jets_sb}\includegraphics[width=0.5\textwidth]{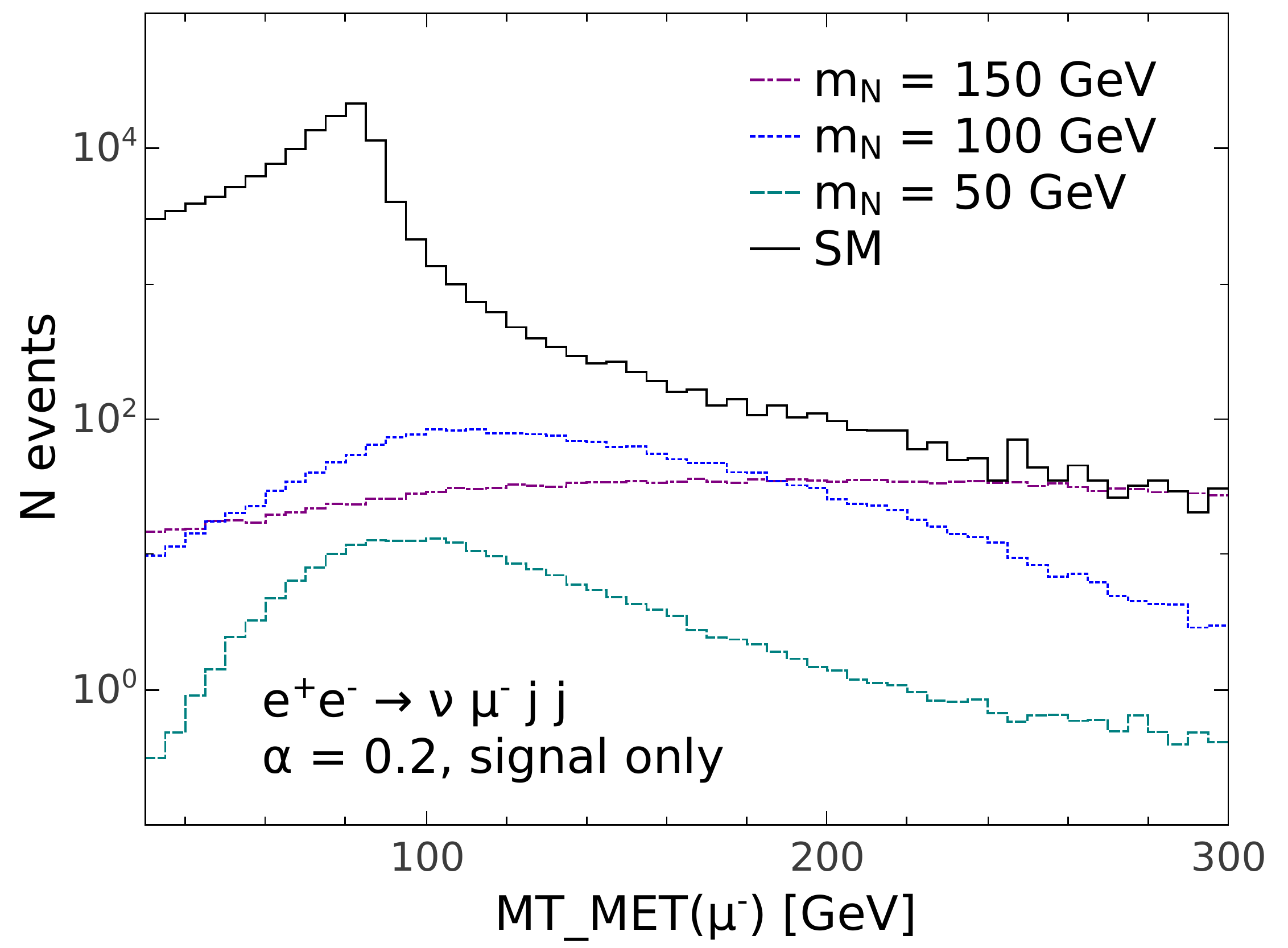}}~
\subfloat[]{\label{fig:AFB_jets_sb}\includegraphics[width=0.5\textwidth]{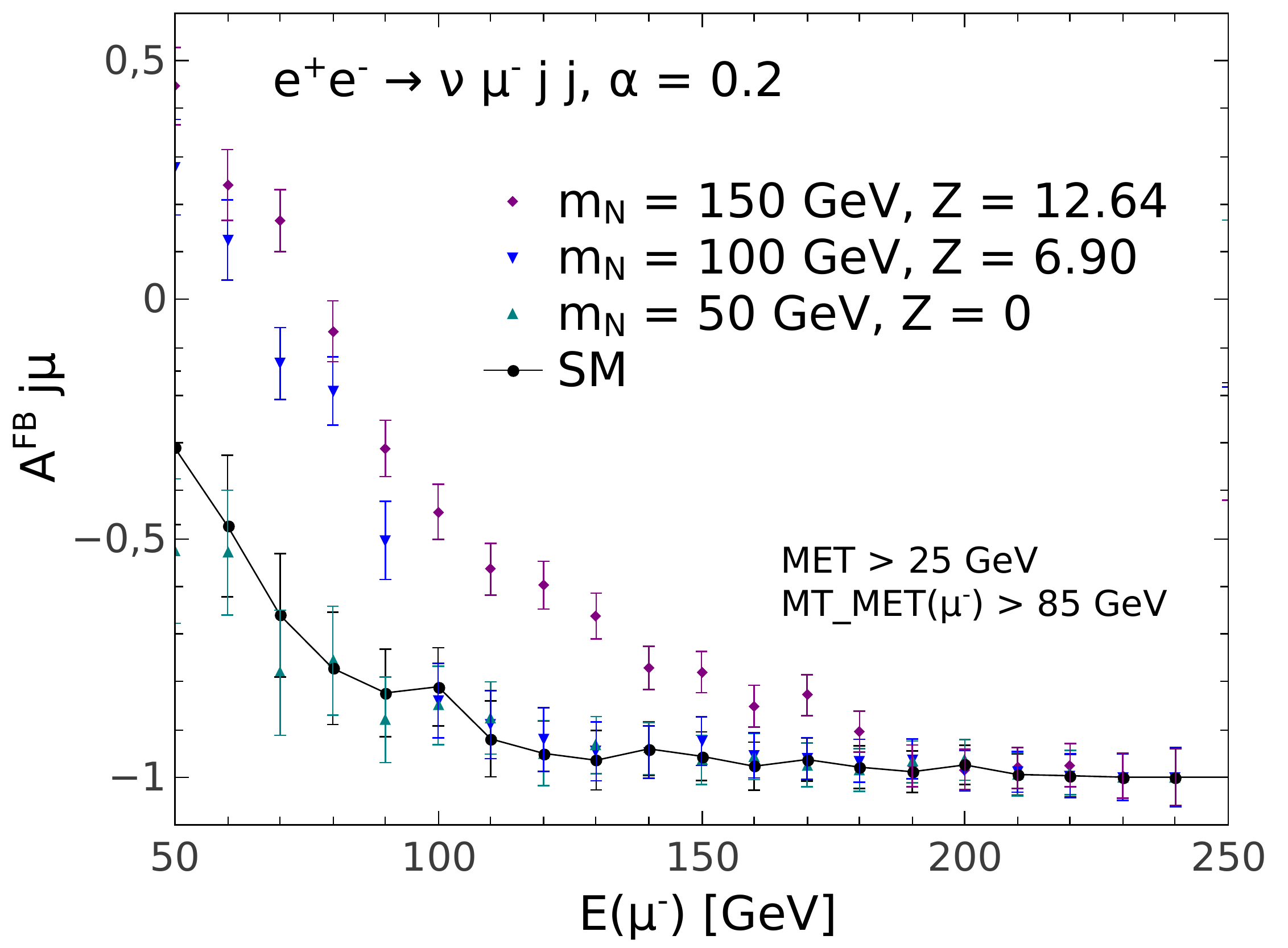}}
\caption{ \label{fig:MTMET_AFB_jets_sb}{Number of events distribution with MT-MET($\mu$) variable (a) and $A_{ \mathrm{j} \mu}^{FB}$ distribution with $E_{\mu}$ for signal plus background (S+B) and SM background-only (B) events for $\alpha=0.2$, $\Lambda=1$TeV, $\sqrt{s}=500$ GeV, $\mathcal{L}=500 ~fb^{-1}$.}}
\end{figure*}

The behavior is different for the SM backgrounds. As we mentioned, the dominant SM contribution to the semi-leptonic process comes from events with pair production of $W$ bosons $ e^{+}e^{-} \to W^{-}  W^{+}$ with $W^{-} \to \nu \mu^{-}$ and $W^{+} \to  \mathrm{j}  \mathrm{j}$. Here the prompt $W^{-}$ and $ W^{+}$ are produced back to back. This gives us muons and jets emerging in opposite directions, and thus $A_{ \mathrm{j} \mu}^{FB}\sim -1$ for the dominant SM background contribution, in events with highly energetic muons, as can be seen in the SM background only (B) events in Fig.\ref{fig:AFB_jets_sb}.

We explored the distribution of the generated signal (S) and background (B) events with several kinematic variables, and select events with missing transverse energy (MET) greater than $25$ GeV to account for the final state neutrino. We also find that the transverse mass of the final muon-neutrino system, calculated in our reconstructed level data as the transverse mass (MT) of the missing transverse energy (MET)- muon system: MT-MET($\mu$) \footnote{The transverse mass variable of the muon-missing transverse energy system is defined as MT-MET($\mu$)= $\sqrt{2 p_T^{\mu} p_T^{miss}[1-cos(\Delta \phi(\vec{ p_T^{\mu}}, \vec{p_T^{miss}})]}$, we use the name given in \texttt{MadAnalysis5}.} can help to separate the signal from the SM background, which peaks at the $m_W$ value, reflecting the fact that the muon-neutrino pair comes from a $W^{-}$. In Fig.\ref{fig:MTMET_jets_sb} we show the number of events distribution with the MT-MET($\mu$) variable for the signals (S) and the SM background (B). We find a cut selecting events with MT-MET($\mu$)>85 GeV keeps more than $70 \%$ of the events for the three $m_N$ signal datasets, while keeping $9\%$ of the SM (B) events. The cutflow is shown in Tab.\ref{tab:cutflow_jets}.  

{\footnotesize{
\begin{table}[t]
 \centering
 \begin{tabular}{ l  l l  l l  l l  l}
 \firsthline 
 \specialrule{.1em}{.05em}{.05em}
$\alpha= 0.2$  & \multicolumn{2}{l}{$m_N=50$ GeV} & \multicolumn{2}{l}{$m_N=100$ GeV} & \multicolumn{2}{l}{$m_N=150$ GeV}  & SM \\
Cuts:  &  S & S+B &  S  & S+B  &  S & S+B & B \\ 
\specialrule{.03em}{.03em}{.03em}
Muon		& $223$ 	&  $134944$  & $2027$ & $136584$ & $2062$ & $136799$ & $135086$ \\  
\specialrule{.02em}{.02em}{.02em} 
 MET  &  $223$(100\%) &   $96046$(71\%) & $2005$(99\%) & $97615$(71\%) & $2048$(99\%)&  $98189$(72\%)&  $96036$(71\%) \\
\specialrule{.02em}{.02em}{.02em}
MT-MET &  $161$(72\%) &  $12140$(9\%) & $1627$(80\%) & $13779$(10\%) & $1755$(85\%)  & $14058$(10\%) & $12051$(9\%) \\
 \specialrule{.03em}{.03em}{.03em}
 $S/\sqrt{(S+B)}~~$  &  $1.46$  &    & $13.86$ &   & $14.8$ &  &  \\
\specialrule{.1em}{.05em}{.05em}
\lasthline 
 \end{tabular}
\caption{\small{$e^{+}e^{-} \to \nu \mu^{-}  \mathrm{j}  \mathrm{j} $ (semi-leptonic). Number of events for signal-only (S), signal and SM background (S+B, generated with interference) and SM background (B), for $\sqrt{s}=500$ GeV, $\mathcal{L}$=500$fb^{-1}$. Muon cut: one muon in final state. MET cut: select MET >25 GeV, MT-MET cut: select MT-MET($\mu$)>85 GeV. Signal significance: $S/\sqrt{(S+B)}$ for event-counting experiment.}}\label{tab:cutflow_jets}
\end{table}
}}

After applying these cuts, we plot the distribution of the asymmetry $A_{ \mathrm{j} \mu}^{FB}$ for muon energy bins $E_{\mu}$ in Fig.\ref{fig:AFB_jets_sb}. The cut applied in MT-MET($\mu$) reduces the number of the dominant SM background events in the lower $E_{\mu}$ bins, giving a less negative value for $A_{ \mathrm{j} \mu}^{FB}$ for the background-only datsaset. We also see that the datasets with signal and background events (S+B) for high $N$ mass values (with more signal-only events than the $m_N=50$ GeV dataset) give even positive $A_{ \mathrm{j} \mu}^{FB}$ values for events with low energy muons, as expected from Fig.\ref{fig:AFB_jets_SS}. 

We find the generated signal plus background (S+B) datasets for $m_N= 100$ GeV and $m_N= 150$ GeV with couplings $\alpha=0.2$ can be separated from the SM background-only (B) dataset with a statistical significance of $6 \sigma$ and $12 \sigma$ respectively. The dataset with $m_N=50$ GeV cannot be separated with the use of the asymmetry, again due to the lower cross section for this signal, as found in Fig.\ref{fig:xs_jets}.

\section{Summary}\label{sec:summary}

The effective field theory extending the standard model with sterile right-handed neutrinos $N_R$ (SMNEFT) parametrizes new high-scale weakly coupled physics in a model independent manner, allowing for a systematic study of their phenomenology in current and future experiments. We consider massive Majorana neutrinos coupled to ordinary matter by dimension 6 effective operators, focusing on a simplified scenario with only one right-handed neutrino added, which provides us with a manageable parameter space to probe. We implement the effective interactions, which include four-fermion operators with the Majorana $N$ field in \texttt{FeynRules}, and make them suitable for simulation with \texttt{MadGraph5} introducing renormalizable UV-interactions with mediators which are integrated out numerically in order to reproduce the low-energy SMNEFT behavior.   
    
We exploit the remarkable angular resolution in future lepton colliders to study the sensitivity of forward-backward asymmetries to discover the possible single production of heavy Majorana neutrinos via $e^{+}e^{-} \to N \nu$, followed by a purely leptonic decay $N \to \mu^{-} \mu^{+} \nu$ or a semi-leptonic decay $N \to \mu^{-} \mathrm{j} \mathrm{j} $, for masses $m_N > 50$ GeV. In this regime, the $N$ production and decays are dominated by scalar and vectorial four-fermion single $N_R$ operators:, $\mathcal{O}_{LNLl}$, $\mathcal{O}_{duNl}$, $\mathcal{O}_{QuNL}$, $\mathcal{O}_{LNQd}$ and $\mathcal{O}_{QNLd}$, as well as the vectorial bosonic $\mathcal{O}_{Nl\phi}$.

This is an alternative analysis to searches using displaced vertices and fat jets, in a higher mass regime, where the $N$ is short-lived but can be found by the angular distribution of its decay products, which depending on their energy can be more or less collimated, leaving a clear imprint on forward-backward asymmetries between them.

By performing a dedicated Monte-Carlo simulation with \texttt{MadGraph5} for the ILC with center of mass energy $\sqrt{s}=500 $~GeV and integrated luminosity $\mathcal{L}=500 ~fb^{-1}$, and implementing a novel analysis with \texttt{MadAnalysis5} using the expert-mode, we have shown that a forward-backward asymmetry between the final muons separation in the pure leptonic decay mode allows us to detect the signal plus background (generated with interference) over the SM background-only events with a sensitivity up to 12$\sigma$ for $m_N=100$ GeV, for effective couplings $\alpha=0.2$ and new physics scale $\Lambda=1$ TeV. In the case of the semi-leptonic decay, we can compare the final muon and higher $p_T$ jet flight directions, and find the highest sensitivity for the $m_N=150$ GeV signal dataset, again with values around 12$\sigma$. 

In summary, we conclude that the use of forward-backward asymmetries in future lepton colliders can probe heavy neutrino new physics in final states with light neutrinos which challenge simpler analyses, with a sensitivity that will eventually allow to put bounds on four-fermion effective operators for a higher $m_N$ mass regime, complementary to proposed displaced vertices searches.

\section{Aknowledgements}
We thank CONICET (Argentina) and PEDECIBA, CSIC, and ANII, under grant PR-FCE-3-2018-148577 (Uruguay) for their financial support.

\appendix


\bibliographystyle{bibstyle.bst}
\bibliography{Bib_N_8_2021}

\end{document}